# $M_1$ - $M^*$ CORRELATION IN GALAXY CLUSTERS

*Dario Trèvese, Giuseppe Cirimele and Benedetto Appodia*
Istituto Astronomico, Università di Roma "La Sapienza",
via G. M. Lancisi 29, I-00161 Roma, Italy

**Abstract**

Photographic F band photometry of a sample of 36 Abell clusters has been used to study the relation between the magnitude $M_1$ of the brightest cluster member and the Schechter function parameter $M^*$. Clusters appear segregated in the $M_1$-$M^*$ plane according to their Rood & Sastry class. We prove on a statistical basis that on average, going from early to late RS classes, $M_1$ becomes brighter while $M^*$ becomes fainter. The result agrees with the predictions of galactic cannibalism models, never confirmed by previous analyses.

## 1. Introduction

A Schechter-like luminosity function (LF) is consistent with the observed galaxy cluster LFs and with a theory of direct hierarchical clustering (Press & Schechter[1], Bond et al.[2]). Galaxy merging has been invoked to modify the Schechter-like shape and reconcile the local LF with faint galaxy count and redshift distribution (see Cavaliere & Menci[3] and refs. therein) According to a galactic cannibalism model (Ostriker and Tremaine[4]), Hausman and Ostriker[5]), brightest cluster members grow at the expense of the other massive galaxies, which are most affected by dynamical friction and this should cause a negative correlation between the magnitude $M_1$ of the brightest cluster member and the characteristic magnitude $M^*$ of a fitting Schechter[6] LF.

Dressler[7] derived an indication that $M_1$ and $M^*$ are negatively correlated from a study of 12 rich clusters. However, subsequent studies of 9 Abell clusters (Lugger[8] (L86)) and 12 clusters (Oegerle and Hoessel[9] (OH89)) found no evidence for any relation between $M_1$ and $M^*$. These results were interpreted as indications against the Dressler[7] claim and the prediction of cannibalism model.

A uniform study of a large sample of nearby galaxy clusters has been undertaken (Flin et al.[10], Trèvese et al.[11] (T92), Trèvese, Cirimele and Flin[12]) to derive their statistical properties. In this paper we report preliminary results of a new analysis of the $M_1$-$M^*$ relation based on a subsample of 36 Abell clusters, more than three times larger than each of the previous samples, from which we obtain a statistically significant evidence of a new type of negative $M_1$-$M^*$ correlation, related to the fact that $M_1$ becomes brighter and $M^*$ fainter going from early to late Rood & Sastry [13] cluster types.

## 2. $M_1$ - $M^*$ correlation

The data were obtained from F-band photographic plates taken with the 48-inch Palomar Schmidt Telescope (Hickson[14]). Plate scanning and data reduction is described in T92. Rel-

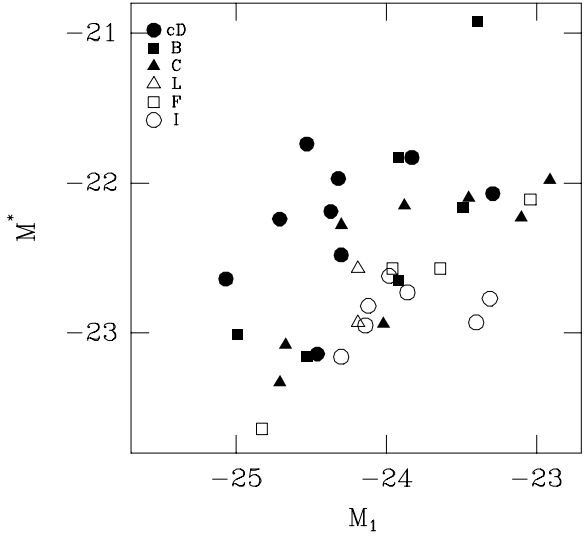 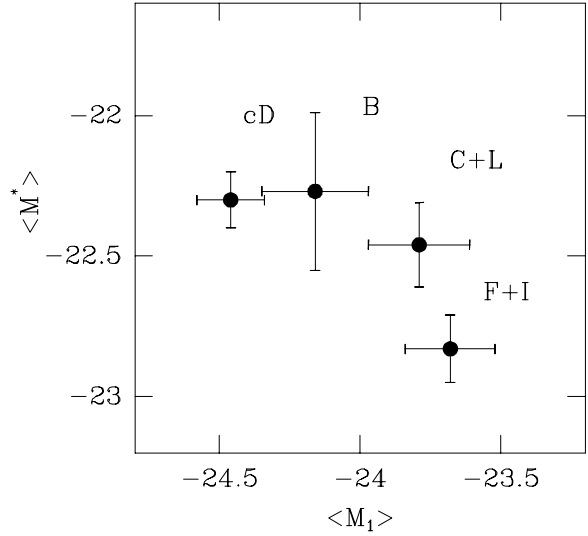

Fig. 1. $M^*$ vs $M_1$ for the clusters sample Different symbols refer to RS classes

Fig. 2. $M^*$ vs $M_1$ averaged over subsamples corresponding to RS classes.

Table 1. The clusters sample

| Abell | z | RS | BM | Abell | z | RS | BM | Abell | z | RS | BM |
|---|---|---|---|---|---|---|---|---|---|---|---|
| A76 | 0.0416 | L | II-III | A655 | 0.1240 | cD | I-II | A1775 | 0.0717 | B | I |
| A147 | 0.0438 | I | III | A656 | 0.136* | cD | I-II | A2028 | 0.0772 | I | II-III |
| A151 | 0.0526 | cD | II | A671 | 0.0494 | C | II-III | A2040 | 0.0456 | C | III |
| A157 | 0.103* | B | II | A779 | 0.0226 | cD | I-II | A2052 | 0.0348 | cD | I-II |
| A260 | 0.0348 | F | II | A1132 | 0.1363 | B | III | A2056 | 0.0763 | C | II-III |
| A278 | 0.0896 | I | III | A1377 | 0.0514 | B | III | A2065 | 0.0721 | C | III |
| A407 | 0.0470 | I | II | A1413 | 0.1427 | cD | I | A2073 | 0.113* | I | III |
| A505 | 0.0543 | cD | I | A1570 | 0.156* | I | II-III | A2096 | 0.108* | C | III |
| A569 | 0.0196 | B | II | A1589 | 0.0718 | C | II-III | A2124 | 0.0654 | cD | I |
| A637 | 0.136* | C |  | A1661 | 0.1671 | F | III | A2593 | 0.0433 | F | II |
| A646 | 0.1303 | I | III | A1689 | 0.1810 | C | II-III | A2657 | 0.0414 | F | III |
| A649 | 0.124* | cD | II | A1700 | 0.119* | L | III | A2670 | 0.0745 | cD | I-II |

The asterisk indicates that z has been estimated from the Abell $z$-$m_{10}$ relation.

ative photometry has been computed for 55 clusters (Trèvese, Cirimele and Flin[12]) and the absolute calibration has been obtained using published photometric data for the 36 clusters listed in table 1 where the redhsifts and RS types are also reported. Color transformation and K-correction from Schneider et. al.[15], have been taken in to account. We estimate an uncertainty in the zero point of the magnitude scale of a few tenth of a magnitude.

The LFs were determined inside circular regions with a fixed radius of $R_3$=1.7/z arc min, corresponding to 3 Mpc for $H_o$=50 $Km\ s^{-1}\ Mpc^{-1}$, $q_o = 1$. The galaxy samples were corrected statistically for the background density and a uniform magnitude limit $m_3 + 3$ was adopted for all clusters. The LFs where then fitted with a Schechter[6] function $\Phi(L)dL = \Phi^* \left(\frac{L}{L^*}\right)^\alpha exp\left(-\frac{L}{L^*}\right) d\left(\frac{L}{L^*}\right)$ ,using a maximum likelihood algorithm. Each LF has been fitted with $\alpha = -1.25$ (Schechter[6]) and $M^*$ as free parameter excluding the first ranked galaxy (see OH89). Figure 1 shows the distribution of the 36 clusters of our sample in the $M_1$-$M^*$ plane. We obtain a positive correlation coefficient $r = 0.48$ and an associated probability $P(> r) = 3 \cdot 10^{-3}$, apparently in contrast with the negative, though non significant, correlation found by Dressler[7]. This could be interpreted as an evidence against the selective depletion of the bright end of the luminosity function, predicted by the galactic cannibalism model of Hausman and Ostriker[5]. However, as can be seen in Figure 1, the clusters are segregated in the $M_1$-$M^*$ plane according to their RS class. Dividing the sample into four groups, corresponding to the RS classes F+I, C+L, B and cD respectively, to collect enough objects in each group. The average values $< M_1 >$ and $< M^* >$ of each group are negatively correlated as seen in Figure 2. The effect is due to the fact that $< M_1 >$ becomes brighter, while $< M^* >$ becomes fainter, going from early to late RS types, consistently with the prediction of the cannibalism model.

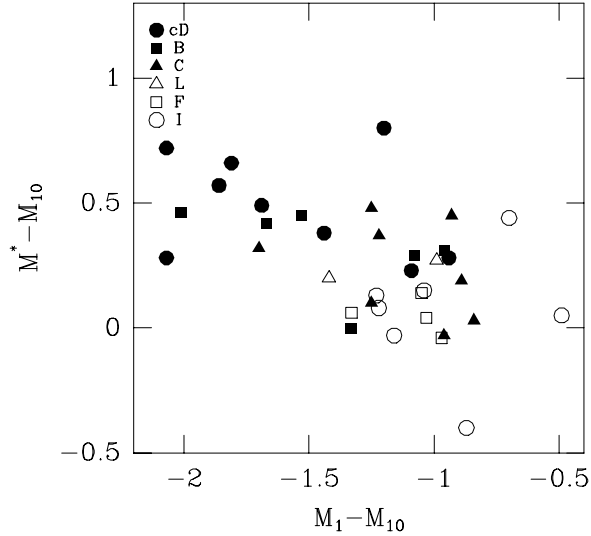

Fig. 3. $M_1 - M_{10}$ versus $M^* - M_{10}$
Different symbols refer to RS classes

The observed positive $M_1$-$M^*$ correlation can be due, at least in part, to uncertainties in the photometric calibration of the plates, since any shift of the magnitude scale affects by the same amount all the galaxy magnitudes. However part of this positive correlation could be intrinsic in nature, e.g. cluster may originate with luminosity functions differing, to a first approximation, by a global shift in absolute magnitude. Assuming as a standard candle the magnitude $M_{10}$ of the tenth brightest member we can plot, see Figure 3, $(M^* - M_{10})$ versus $(M_1 - M_{10})$, which are independent of the calibration uncertainties and appear negatively correlated. Also a partial correlation analysis indicates the same effect. The global correlation coefficients between $M_1$, $M^*$ and $M_{10}$ are all positive : $r_{1,*}=0.54$ , $r_{1,10}=0.82$ and $r_{*,10}=0.87$. The partial correlation: $r_{1,*;10} = (r_{1,*} - r_{*,10} \cdot r_{1,10})/((1 - r_{*,10}^2) \cdot (1 - r_{1,10}^2))^{\frac{1}{2}}$ which represents the 'intrinsic' correlation between $M_1$ and $M^*$ is $r_{1,*;10} = -0.61$ with an associated probability $P(> |r|) = 8 \cdot 10^{-5}$.

The effect is statistically significant, thus providing a new constraint for any model of cluster formation and evolution.

## 3. Conclusions

We have determined the luminosity functions of a sample of 36 Abell clusters and fitted them with Schechter-like profiles, assuming a canonical $\alpha = -1.25$, and we find that:

- On average, going from early to late Rood & Sastry types, the magnitude $M_1$ of the bright cluster member becomes brighter, while the characteristic magnitude $M^*$ is fainter. The effect is statistically significant, providing a new constraint for theories of cluster formation and evolution.

- Including in the study also the magnitude $M_{10}$, assumed as a standard candle, a partial correlation analysis confirms a negative intrinsic correlation between $M_1$ and $M^*$.

- These results support the cannibalism model of Hausman & Ostriker, at variance with previous analyses.

Part of the positive $M_1$ - $M^*$ correlation is caused by uncertainties in the absolute photometry while part could be intrinsic in nature.